\documentclass[letterpaper,11pt]{article}
\usepackage{authblk}
\usepackage{amstext,amsmath,amssymb,amsfonts,bbm, amsthm}
\usepackage[latin1]{inputenc}
\usepackage{array,graphicx}
\usepackage{hyperref}
\usepackage{tocvsec2}
\usepackage[top=2cm, bottom=2cm, left=1.8cm, right=1.8cm]{geometry}
\usepackage{color}
\usepackage{lastpage}
\usepackage{tabularx}
\usepackage[table]{xcolor}
\usepackage{enumitem}
\usepackage{etaremune}
\usepackage{multirow}
\usepackage{pdfpages}
\usepackage{setspace}

\usepackage{pdfpages}

\usepackage{booktabs}
\usepackage{pifont}

\usepackage{eurosym}

\newcommand\ci{\mathbin{\perp\kern-9.4pt\perp\,}}

\usepackage{setspace}

\begin{document}

\clearpage

\title{Mind before matter: reversing the arrow of fundamentality}
\author{Markus P. M\"uller}
\affil{\footnotesize Institute for Quantum Optics and Quantum Information, Austrian Academy of Sciences, Boltzmanngasse 3, A-1090 Vienna\\
Perimeter Institute for Theoretical Physics, Waterloo, ON N2L 2Y5, Canada}

\date{}

\maketitle

\begin{abstract}
In this contribution to FQXi's essay contest 2018, I suggest that it is sometimes a step forward to reverse our intuition on ``what is fundamental'', a move that is somewhat reminiscent of the idea of noncommutative geometry. I argue that some foundational conceptual problems in physics and related fields motivate us to attempt such a reversal of perspective, and to take seriously the idea that an information-theoretic notion of observer (``mind'') could in some sense be more fundamental than our intuitive idea of a physical world (``matter''). I sketch what such an approach could look like, and why it would complement but not contradict the view that the material world is the cause of our experience.	
\end{abstract}

\section{Prequel}
There was this young man who had tried to make Nadine drink just a tiny jar of water. It was a strange game they were playing, every day, day by day: the four-year old who wouldn't drink versus the nineteen-year old who knew that her life depended on it. Little stubborn girl versus clumsy determined teenager.

That day, he lost the game again.

Sad and worried, he gave up. He lifted Nadine from her child's chair and sat her on the ground, where she could do what she liked most: play and explore.

Nadine was always on the brink of dehydration, but she was a true discoverer. Almost blind and multiply challenged, she could move only one arm, which meant that she was crawling on the floor in a circle. But what a beautiful circle it was! All smiling and her eyes lit up, she was rolling her ball, touching and moving her toy bricks, and discovering her big little world with grace and determination.

There was something that began to dawn on him. Nadine seemed like a prisoner of her body, and her circles and limitations a perfect symbol for the brutal power of the material world over her self. Physics tells us that all there is supervenes on the atomic building blocks of this one, fundamental world of cruel concreteness.

Or does physics, really? Could the light in Nadine's eyes convey a message for us that things are truly different?

What if we got this all wrong?

\section{The orthodox view...}
Many of us are skeptics, and so am I. We reject ideas like astrology, omnipotent gods, or the afterlife simply because there is no convincing evidence for any of those things. We know how easily we can deceive ourselves and how often we err, which is why science is our method of choice.

What is it that makes science trustworthy? Philosophers have long been arguing about how to best define the scientific method and how to delineate it from pseudoscience (see e.g.~\cite[Section 4]{StanfordEncycl} for an overview), but there seems to be widespread consensus that features of self-correction play an important role. We try to adapt our views to new evidence, we reproduce our experiments many times, and we test our findings against those of others. The results of this approach, refined via mathematics and statistics, technological craftsmanship, and philosophical reflection, represent the closest to objective knowledge that we have.

Cherishing reliable objectivity, we can easily understand why the scientific community at large promotes what I call here \emph{the orthodox view}, and we should in fact be glad that it does. According to the orthodox view, there is a single, material universe that evolves in time according to physical laws, and this is fundamentally all there is to say. In particular, ``observers'' or ``agents'' play no foundational role whatsoever. The question of consciousness is deliberately ignored, and the lessons of the Copernican Revolution~\cite{Kuhn} are extrapolated and promoted to a paradigm: the earth is not the center of the universe, humans are not central to physics, hence the ``self'' should not play any distinguished role in science.

We should be more than happy that the orthodox view dominates: banishing subjectivity, religious authority, and unverifiable spiritual claims was arguably a crucial prerequisite for progress and enlightenment. Without its development, we would not be able to give our children vaccination or antibiotics, and, more importantly, we would not have any good reason to tell them that they need not be afraid of ghosts. Moreover, most contemporary attempts to go beyond this view are pseudo-scientific or at least highly controversial, such as the idea that ``consciousness collapses the wave function'' or the proliferation of the anthropic principle in the absence of predictive power of a theory.

Yet, there are indications that the orthodox view is incomplete. One such indication is the hard problem of consciousness. As Chalmers puts it~\cite{Chalmers}, \emph{``It is widely agreed that experience arises from a physical basis, but we have no good explanation of why and how it so arises. Why should physical processing give rise to a rich inner life at all? It seems objectively unreasonable that it should, and yet it does.''} There are arguably good reasons to conjecture that the orthodox view might be unable in principle to provide a basis for solving this problem. But the focus of this essay is not on consciousness: as I will argue below, there are several problems \emph{in and around physics} that point to a systematic deficiency of the orthodox view. These problems can guide our attempts in exploring alternatives to the orthodox perspective.

\begin{figure}[!hbt]
\begin{center}
\includegraphics[angle=0, width=12cm]{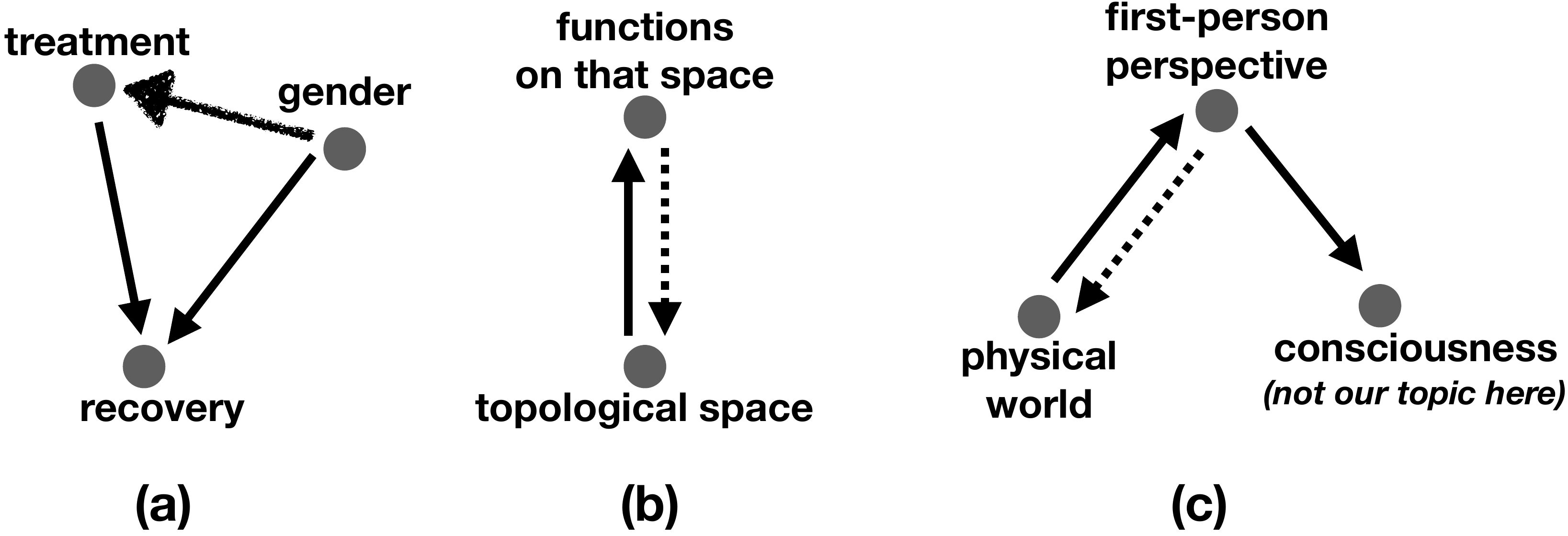}
\caption{\footnotesize \textbf{(a)} In the theory of causality, directed acyclic graphs (DAGs) are used to represent the causal structure of a set of random variables. In this example, we would have a medical treatment that influences patients' recovery, but also gender impacting recovery. If the hand-drawn arrow is present, then we have an additional influence of gender on the willingness of receiving treatment, which leads to counterintuitive effects like Simpson's paradox; see e.g.\ the book by Pearl~\cite{Pearl}. We are here borrowing this graphical notation for representing \emph{fundamentality} or \emph{supervenience} --- in a nutshell, we draw an arrow from A to B if A ``comes before'' B in some well-defined sense, i.e.\ if A is more fundamental than B. \textbf{(b)} If we are given a topological space, then we can define the (real) functions on that space. In this mathematical context, we would therefore intuitively say that the functions supervene on the space, and that we should draw an arrow from the (more fundamental) topological space to the (derived) algebra of functions on it, which would lead to the solid arrow. But, as explained in the main text, noncommutative geometry reverses this, leading to the dotted arrow instead (erasing the solid arrow), and does so with benefits. \textbf{(c)} In what I call the ``orthodox view'', one would draw the two solid arrows: the physical world is the fundamental basis on which everything else supervenes. Let us introduce an abstract notion of ``first-person perspective'' as, roughly speaking, the information-theoretic content of an observer's brain. Then, clearly, the orthodox perspective says that this is a property of the material world, and then consciousness is somehow supervening on that abstract first-person perspective. (On the other hand, \emph{dualists} would probably erase at least one of the arrows.) The hard problem of consciousness is then to understand how these arrows come about, i.e.\ what sort of ``causation'', logical implication, or supervenience they are supposed to represent. But instead of asking this question, my proposal here is to do something similar as in (b): reverse the arrow of fundamentality.}
\label{fig_dags}
\end{center}
\end{figure}

Given the benefits of the orthodox view, I am certainly not suggesting to drop it. Instead, I propose to modify it in a way that is consistent with the fundamental tenets of science, while keeping it fully intact in the familiar regime of physics. In a nutshell, my suggestion will be to ``reverse the arrow of fundamentality'': it is not the external world that is ultimately fundamental and the self that supervenes on it, but rather the other way around in a specific sense (see Figure~\ref{fig_dags} for an illustration of the ``arrow'' terminology).

Reversing the arrow in this sense is not unprecedented in science. Quite on the contrary: for example, \emph{noncommutative geometry}~\cite{Lizzi} can be seen as such a reversal, and it will be instructive for what follows to examine this example in a bit more detail. Consider a topological space\footnote{I am omitting some mathematical details here to keep the presentation accessible.}, called $X$. Once we have this space, we can look at the continuous real functions on it, denoted $C(X)$. This set of functions has an important property called \emph{commutativity}: the order of multiplication doesn't matter, i.e.\ we have $fg=gf$ for any two such functions. It seems completely obvious that $X$ is more fundamental than $C(X)$, in the sense that it ``comes first'' in the logical architecture of mathematical objects (see also the solid arrow in Figure~\ref{fig_dags}(b)). However, it turns out that this logical path can be reversed in some sense: if we know the set of functions $C(X)$ and their algebraic properties, we can in principle reconstruct the underlying topological space $X$.

Noncommutative geometry takes this observation seriously, and uses it to generalize the notion of a ``space''. Namely, instead of considering a set of functions, it instead starts with an ``algebra'' $A$ of objects which need not be commutative, i.e.\ it is allowed that $fg\neq gf$. It then uses the mathematical methods that led from $C(X)$ back to the space $X$, and leads us from $A$ to ``something''. Is that ``something'' an underlying space for $A$? Well, not quite --- it is a \emph{noncommutative space}. It is similar to ordinary spaces in some respects, but different in others. In particular, one hopes that it can represent the sort of ``quantum spacetime'' that one expects to find in physics in the realm of quantum gravity. If successful, the mathematical strategy of noncommutative geometry would then attain an attractive physical interpretation: it would mean that quantum theory (and its algebra of operators) is more fundamental, and (some generalized notion of) spacetime supervenes on it. This is a reversal of the ``usual'' arrow of fundamentality.

Reversing the arrow may work in the context of noncommutative geometry, but why should we adopt this strategy for the ``self'' versus the ``physical world''? How could this even work? Let us look at some problems of physics for guidance and motivation.

\section{... and its limitations in a world that is large, technologically interesting, or of the quantum kind}
\label{SecLimitations}
The orthodox view with its notion of fundamental physical world works perfectly fine --- in a certain regime. Namely, it applies perfectly to a rather small universe which does not contain too powerful technology and in which we can ignore quantum theory most of the time. It is in this regime where we can easily implement the solid arrow at the left of Figure~\ref{fig_dags}(c): starting from the laws of that physical world, and our ability to predict its evolution, we can compute (probabilistic) predictions for the first-person perspective. That is, we can use our physical theories to \emph{predict what we, as observers, will see in certain situations}. For example, if we pick up a stone, lift it with our hand, and release it, we can predict that we will (much) more likely see it subsequently fall down than see it fall up, using the laws of mechanics. This is crucial because this is what allows us to test our theories, comparing predictions with actual observations.

Nevertheless, this way of thinking leads to problems and paradoxes if our universe is \emph{very large}, like in certain scenarios involving eternal inflation. In this case, there are cosmological models in which the universe is full of very improbable but (due to its size) numerous thermodynamic fluctuations~\cite{Albrecht,AlbrechtSorbo}. What, then, tells us that \emph{we} are not one of those fluctuations (the infamous ``Boltzmann brains'') who have just come into existence by a combinatorial accident? All the memories of our past lives in an ordered, planetary, low-entropic environment would then be mere illusions, and in the next moment, we would make a very scary and unexpected experience before evaporating forever in the midst of nowhere. Shouldn't we assign much higher probability to such a shocking experience than to an ordinary continuation of our lives if our cosmological models tell us that the universe contains much more Boltzmann brains than ordinary brains?

Note that I am not claiming that this is the right way to think about the problem; I am simply pointing out \emph{that it is a problem in the first place}, one that makes cosmologists wonder and argue. The orthodox view itself does not tell us (at least not directly) how to deal with questions like this because we have no idea how we should reduce this question to a question about the physical world.

We need not believe in eternal inflation or turn to cosmology to run into problems of this kind; we can create our own ``Boltzmann brain problems'' with technology. For example, imagine that some scientists put you to sleep and scan your brain in great detail (while unfortunately destroying it), only to create a near-perfect computer simulation~\cite{Bostrom} of your brain, connected to a simulated body in some simulated environment. Moreover, suppose that the scientists create a large number of slightly different copies, running on different types of computers, possibly delayed in time. Would you ``wake up'' in a simulation? If so, in which one? Shortly before the experiment, what probability should you assign to finding yourself in any given simulation? It seems that physics must be silent about this question in principle, which is odd: isn't the very essence of physics that it tells us what we will see next given what we have seen before?

Or is this demand misguided, and the essence of physics is ``to tell us what is really going on in the world''? Not if we live in a quantum world. Contextuality~\cite{Peres} tells us that it is impossible to assign truth values to all propositions (represented by projection operators) such that a measurement simply reveals the corresponding value; thus, in a nutshell, it is \emph{inconsistent} to assume that the world has (only) well-defined properties that we are able to uncover by inspection or measurement. This insight can be cast into many different precise mathematical statements, from Bell's theorem~\cite{Bell1964,Bell1966} to no-go theorems about ``facts of the world''~\cite{Brukner,FrauchigerRenner}. The upshot is that quantum physics only tells us what results to expect with which probability if we decide to perform a certain measurement. In this sense, quantum physics, the most accurate and successful theory we have ever had, talks directly about \emph{what we see} (conditional on how we observe) and not \emph{what there is} (in a naive sense). From an orthodox perspective, this is highly surprising.

It would not be so surprising if we reversed the arrow in Figure~\ref{fig_dags}(c), and considered an abstract, information-theoretic notion of first-person perspective (not consciousness!) as more fundamental. Then the physical world would be an emergent, less fundamental notion, and we should expect our most fundamental theories to talk about what is seen and not what there is. Then we should also be prepared to find phenomena comparable to those in noncommutative geometry: while the latter leads to ``something close to ordinary space, but not quite'', we should analogously expect to obtain ``something close to an ordinary world, but not quite''. Which in some sense we do --- we live in a quantum world.

But if we take this idea of ``reversing the arrow'' seriously, how can we concretely make this work?

\section{From mind to matter...}
In Ref.~\cite{Draft} (see~\cite{DraftShort} for a summary), I have constructed a ``proof of principle'' theory that ``reverses the arrow'' in the sense explained above: it starts with the ``self'', and shows that an emergent notion of ``world'' follows. This is not the place to go into all the details, so let me simply give a very brief overview.

The starting point is to formalize an information-theoretic notion of ``your state'' at a given moment (as mentioned in Figure~\ref{fig_dags}(c), we do not intend to talk about ``consciousness'' or ``qualia'' here but aim for a technical notion). Think of everything that you, as an observer, perceive and remember at some given moment --- something like a raw dump of all the data in your brain. We will denote this raw data by a finite string of bits, something like $x=011010$ (just typically much longer). When we write down such a string, we assume that it makes sense to talk about ``being in that state'', in the sense that there is a corresponding first-person perspective, i.e.\ a notion of ``experiencing to be in that state and not another one''. In other words, we assume that there is some ``mental oomph'' that is described by any given string of bits, in a similar way as we typically ascribe some corresponding ``material oomph'' to whatever is described by giving the location of the positions and velocities of particles (or properties of a quantum field) in physics.

We assume that ``being an observer'' means to be in some state $x$ at any given moment, and then to be in another state $y$ in the subsequent moment\footnote{Here, ``moment'' does not refer to some externally given time, but to an integer labelling of the subjective states.}. We typically think that $y$ is determined by physics, that is, by the external world. For example, if $x$ describes that I see a tile fall from the roof of my house, then $y$ will typically encode that I see the tile fall further down (or hit the ground) because that is what happens in the world, and my brain is part of that world. But if we do not presuppose the existence of a ``world'', then we cannot resort to that argumentation. Instead, we need a ``law'' that acts directly on the observer states, telling us what state to expect next, without assuming that it derives from some physical universe.

This is done by the following postulate\footnote{As explained in~\cite{Draft}, it would be more natural to formulate this postulate in terms of a \emph{Markovian} probability measure, one for which the (probability of the) next state only depends on the current state and not on all previous ones. However, finding such a formulation and exploring its properties is mathematically much more challenging; it is currently an open problem.}:

\textit{\textbf{Postulate 1.} Being an observer means to be in some state $x_1$ first, then in some state $x_2$, and so on. The probability (chance) of being in state $y$ next, after having been in states $x_1,\ldots,x_n$, is given by conditional algorithmic probability $\mathbf{P}(y|x_1,\ldots,x_n)$, i.e.\ $\mathbf{P}(x_1,\ldots,x_n,y)/\mathbf{P}(x_1,\ldots,x_n)$.}

What is algorithmic probability? In a nutshell\footnote{Interested readers should look at~\cite{LiVitanyi} and~\cite{Draft} for the correct mathematical definitions.}, it is a probability distribution that \emph{favors compressibility}. The probability $\mathbf{P}(x_1,\ldots,x_n)$ is roughly $2^{-L}$, where $L$ is the length of the shortest computer program (formalized by a version of universal monotone Turing machines) that produces first the output $x_1$, then the output $x_2$, and so on, until $x_n$ (and then possibly more). Consequently, if $\mathbf{P}(y|x_1,\ldots,x_n)$ is large, then this means that short programs tend to output $y$ after having output $x_1,\ldots,x_n$ --- in other words, that it is somehow (algorithmically) ``natural to guess'' that $y$ comes next.

 In~\cite{Draft}, I give three different conceptual and structural-mathematical reasons for postulating this distribution and not another one. Without going into this argumentation, the most obvious indication of the relevance of algorithmic probability $\mathbf{P}$ comes from ``Solomonoff induction''~\cite{LiVitanyi}: in computer science and artificial intelligence~\cite{Hutter}, $\mathbf{P}$ is shown to be an efficient tool for predicting future observations, under some computability assumptions that are satisfied in physics according to some version of the Church-Turing thesis. Postulate 1 then claims that $\mathbf{P}$ does not only predict the future, but in fact determines it.

What are the consequences of Postulate 1? At this point, the  mathematical tools of algorithmic information theory become relevant, leading us to some quite surprising predictions. One such predictions is what I call the ``principle of persistent regularities'': if there has been a computable regularity in all previous observations (say, by mere chance), then there is a high probability that this regularity will be present also in future observations. In more detail, define a ``computable test''\footnote{This is a special case of the definition in~\cite{Draft}. See also~\cite{Draft} for issues related to Goodman's New Riddle of Induction.} as a computable function that assigns to any bit string $x$ some $f(x)$ which is either ``yes'' ($1$) or ``no'' ($0$). Then one can prove the following:

\emph{\textbf{Theorem 1.} Consider the conditional probability that $f$ will yield ``no'' next, if it has given the answer ``yes'' on all previous observer states; i.e.\ $\mathbf{P}(0|1^n):=\mathbf{P}(f(y)=0|f(x_1)=\ldots=f(x_n)=1)$. Then $\lim_{n\to\infty}\mathbf{P}(0|1^n)=0$, and the convergence is rapid, since $\sum_{n\in\mathbb{N}}\mathbf{P}(0|1^n)$ converges. That is, if $n$ is large, then the answer will probably be ``yes'' next, too.}
 
As a simple example, think of a computer program $f$ that checks whether $x$ corresponds to a ``brain dump'' that is typical for an observer in a planet-like environment. If that check gave the answer ``yes'' to all previous observer states (and there were enough of those), then it would give ``yes'' with high probability also to future observer states. But this resolves the Boltzmann brain problem, regardless of any assumptions on cosmology (for pages of painstaking details, see~\cite{Draft}). Another way to see this is that  Solomonoff induction would never make an observer assign significant probability to a shocking Boltzmann brain experience as described in Section~\ref{SecLimitations}, and according to Postulate 1, Solomonoff induction is correct by definition.

It is this tendency to ``stabilize regularities'' that ultimately leads to an emergent notion of external world, as mentioned in the beginning of this section. To understand the significance of the following theorem, it makes sense to first read its illustration and interpretation in Figure~\ref{fig_world} below.

\emph{\textbf{Theorem 2.} Consider any computable probabilistic process which has description length $L$ on a universal computer; we say that this process is \emph{simple} if $L$ is small. Suppose that this process generates a sequence of bit strings $x_1,x_2,\ldots$ as outputs, with probability $\mu(x_1,\ldots,x_n)$. Then, with $\mathbf{P}$-probability of at least $2^{-L}$, we have $\mathbf{P}(y|x_1,\ldots,x_n)\longrightarrow \mu(y|x_1,\ldots,x_n)$ for $n\to\infty$, i.e.\ this (simple) computable probabilistic process will asymptotically yield a perfect probabilistic description of the observer's state transitions.}
\begin{figure}[!hbt]
\begin{center}
\includegraphics[angle=0, width=8cm]{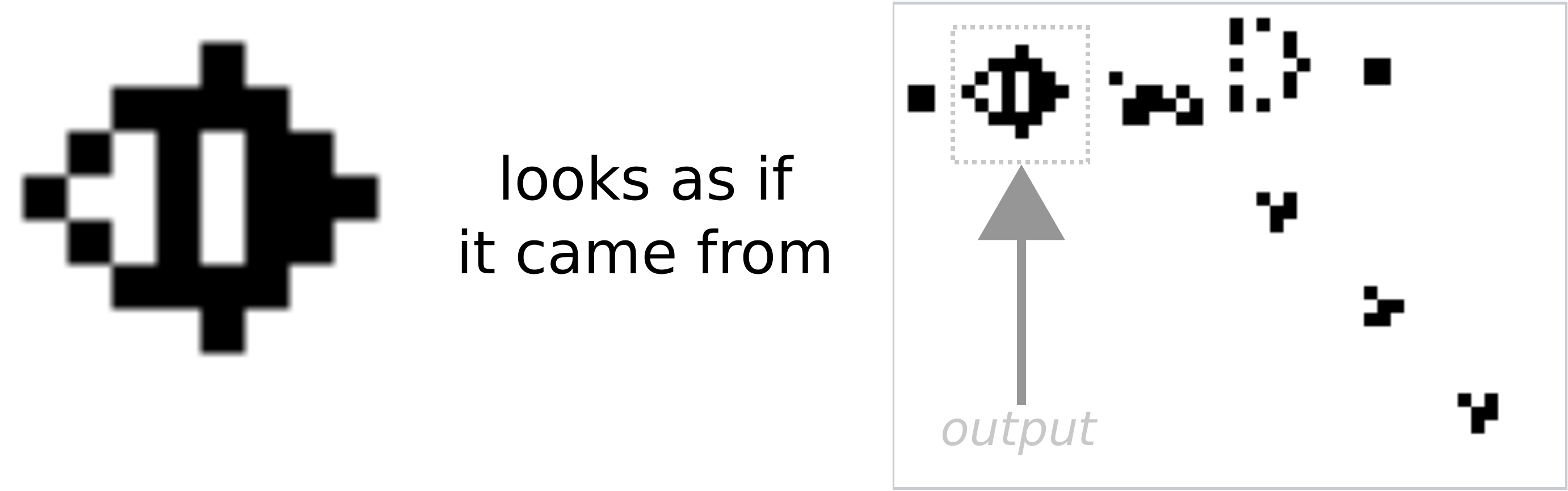}
\caption{\footnotesize The left-hand side symbolizes a bit string $x$, representing the state of an observer, and the right-hand side represents a probabilistic computable process (here actually deterministic: an instance of Conway's Game of Life~\cite{GameOfLife}), together with a rule to ``pick out'' some random variable from the process (playing the role of an ``output''; instead of a distinguished tape as for a Turing machine, it is here some computable ``locator function'' that defines the output). Postulate 1 acts directly on the left-hand side: it says that algorithmic probability determines how the observer's state changes over time. But Theorem 2 shows that, \emph{as a consequence}, after having run through many states $x_1,\ldots,x_n$, the observer's state will ``look as if'' the observer was actually the output part of some such computational process, in the sense that the probabilities $\mathbf{P}$ of observer state changes on the left will be equal to the marginal probabilities of some part of the process (the output) on the right. For example, if one glider in Gosper's glider gun is going to hit the observer in the process in the next step, then this is reflected on the left by a corresponding change of state of the observer. In this sense, the (rest of) that computational process is something like an ``external world'': it is not directly accessible to the observer, but correlated with the observer's future states. It is a ``convenient fiction'' to predict the future --- namely, an emergent notion of a physical universe that admits mechanistic causal explanations.}
\label{fig_world}
\end{center}
\end{figure}

Therefore, we obtain a prediction that seems consistent with the facts: observers will, with high probability, asymptotically be in states that look as if they were part of a larger computable, probabilistic process --- an ``external world''. The simpler the world (i.e.\ the smaller the $L$) the more probable that it emerges.

There would be much more to say about the consequences of Postulate 1: namely, that we also get an emergent notion of objective reality among \emph{several observers}, that the emergent world doesn't have to look like a typical computation on our desktop computers, that we expect to find some features (but not necessarily all properties of) quantum theory, and that there are surprising novel predictions like ``probabilistic zombies''. But for these and other aspects, I refer the reader to~\cite{Draft}. Instead, let us discuss what a theory of the kind described above would imply for the question of fundamentality and causality.

\section{... and back from matter to mind: a strange loop of fundamentality}
There is something deeply puzzling about the above: if the mind is more fundamental than the world, then what about our familiar notion of causality? For example, don't we have a coherent explanation for the kind of technical, information-theoretic content of our brain as a \emph{result} of the laws of physics? The formation of the solar system, the genesis of the first life forms (despite our missing knowledge about the details of this event), and the subsequent process of Darwinian evolution are explanatory triumphs of science that allow us to understand perfectly well why there are functional brains in the first place, and why they roughly have the informational structure they do. Does the theory above claim that all this is wrong?

The answer is a clear ``no'' --- this standard explanation is still available and perfectly valid. The catch is that there are now two possible and mutually compatible perspectives to take. This can be seen by example of Figure~\ref{fig_world}: on the one hand, we can argue directly via an observer's state, as on the left-hand side. Postulate 1 tells us that algorithmic probability determines what happens to an observer, and the right-hand side can be seen as a consequence of this: the properties of algorithmic probability imply that some notion of external world emerges. But, \emph{by the very definition of what this means}, this emergent external world gives an excellent description of what happens to the observer state, since its output configuration evolves under the same probabilities as that state. For example, if (on the right-hand side) a glider collides with the observer's part of the grid, then (on the left-hand side) there will be a corresponding state change of the observer. It is therefore consistent, for all (not only practical) purposes, to regard the collision with the glider as the \emph{cause} of that state change.

In other words, since this emergent world corresponds to a simple algorithm which represents an excellent compression of the observer's probabilistic state changes, we can regard its functioning as the background ontological structure that gives rise to what the observer sees. Thus, we can use it to obtain algorithmic, causal, or ``mechanistic'' explanations for the observer's states (including evolutionary explanations), but we may want to keep in mind that this background algorithm is ultimately itself not fundamental.

Given these two possible perspectives, it becomes somewhat unclear how we should ``draw the arrow'' in Figure~\ref{fig_dags}(c): in some sense, we have ``reversed the arrow'' by declaring the first-person perspective to be more fundamental than the physical world. On the other hand, in the resulting worldview, the emergent external world can nevertheless consistently be viewed as the sole mechanistic basis, and thus \emph{cause} in a physical sense, of that first-person perspective. In the end, we arrive at a picture that resembles John A.\ Wheeler's picture of a universe observing itself~\cite{Wheeler}: a ``strange loop'' of mind and matter, subsequently giving rise to each other, and supervening on the respective other, in conceptually slightly different ways.

In summary, we learn from this approach that an ultimate notion of fundamentality may have a very subtle structure. On the one hand, ``reversing the arrow'', i.e.\ turning our idea of the direction of supervenience upside down, can lead to novel insights that are not otherwise available, as the examples of noncommutative geometry and the approach sketched above have shown. On the other hand, the resulting worldview can exhibit surprising features that undermine our intuitive ideas about fundamentality, including a disidentification with causality, perhaps confirming views like Bertrand Russell's skepticism towards the latter. These surprises may well be relevant for approaching some notorious open questions in the foundations of physics.

\section{Sequel}
Almost twenty years after my struggles with Nadine's drinking habits, I was very happy to find that the material world had not been able to exert as much brutality on her as everybody had first thought. My teenager self had been told that Nadine has only a few years left to live; instead, the last birthday she was able to celebrate (without a drink I suppose) was her twentieth. Sixteen more years of exploration!

It is the orthodox methodology of science that allows us to help people like Nadine --- to diagnose illnesses, to understand the underlying mechanisms, to design medication that helps reliably. We should be proud to have come so far. It is the same science reminding us that the light in Nadine's eyes is telling us literally nothing that would in itself justify the idea that our orthodox perspective is limited.

But maybe physics and mathematics will do, at some point.

If the ideas above contain a grain of truth, then the mind may ultimately be more fundamental than the world, in some specific sense. And this may allow us to approach questions like Chalmers' with completely new ideas in our heads, and to look at Nadine's struggles with a new sense of hope in our hearts.

\textit{--- Dedicated to Nadine, and all the other fearless stubborn explorers out there.}


\begin{thebibliography}{99}

\bibitem{StanfordEncycl}
S.\ O.\ Hansson, \emph{Science and Pseudo-Science}, The Stanford Encyclopedia of Philosophy (Summer 2017 Edition), Edward N.\ Zalta (ed.), URL=\url{https://plato.stanford.edu/archives/sum2017/entries/pseudo-science/}.

\bibitem{Kuhn}
T.\ S.\ Kuhn, \emph{The Copernican Revolution}, Harvard University Press, Cambridge, Massachusetts, 1957.

\bibitem{Chalmers}
D.\ Chalmers, \emph{Facing up the problem of consciousness}, in: D.\ Chalmers, \emph{The character of consciousness}, Oxford University Press, 2010.

\bibitem{Pearl}
J.\ Pearl, \emph{Causality -- models, reasoning, and inference}, Cambridge University Press, New York, 2009.

\bibitem{Lizzi}
F.\ Lizzi, \emph{The Structure of Spacetime and Noncommutative Geometry}. In: Geometry, Topology, QFT and Cosmology, 28-30 Maggio 2008, Paris, France (2009).

\bibitem{Albrecht}
A.\ Albrecht, \emph{Cosmic Inflation and the Arrow of Time}, in \emph{Science and Ultimate Reality: From Quantum to Cosmos}, honoring John Wheeler's 90th birthday, J.\ D.\ Barrow, P.\ C.\ W.\ Davies, and C.\ L.\ Harper (eds.), Cambridge University Press, 2004.

\bibitem{AlbrechtSorbo}
A.\ Albrecht and L.\ Sorbo, \emph{Can the universe afford inflation?}, Phys.\ Rev.\ D \textbf{70}, 063528 (2004).

\bibitem{Bostrom}
N.\ Bostrom, \emph{Are you living in a computer simulation?}, Philosophical Quarterly \textbf{53}(211), 243--255 (2003).

\bibitem{Peres}
A.\ Peres, \emph{Quantum Theory: Concepts and Methods}, Kluwer Academic Publishers, New York, 2002.

\bibitem{Bell1964}
J.\ S.\ Bell, \emph{On the Einstein Podolsky Rosen Paradox}, Physics \textbf{1}(3), 195--200 (1964).

\bibitem{Bell1966}
J.\ S.\ Bell, \emph{On the problem of hidden variables in quantum mechanics}, Rev.\ Mod.\ Phys.\ \textbf{38}(3), 447--452 (1966).

\bibitem{Brukner}
\v{C}.\ Brukner, \emph{On the quantum measurement problem}, in \emph{Quantum (Un)Speakables II --- Half a Century of Bell's Theorem}, R.\ Bertlmann and A.\ Zeilinger (ed.), Springer International Publishing Switzerland, 2017.

\bibitem{FrauchigerRenner}
D.\ Frauchiger and R.\ Renner, \emph{Quantum theory cannot consistently describe the use of itself}, Nat.\ Comm.\ \textbf{9}, 3711 (2018).

\bibitem{Draft}
M.\ P.\ M\"uller, arXiv:1712.01826.

\bibitem{DraftShort}
M.\ P.\ M\"uller, arXiv:1712.01816.

\bibitem{LiVitanyi}
M.\ Li and P.\ Vit\'anyi, \emph{An Introduction to Kolmogorov Complexity and Its Applications}, Springer, 1997.

\bibitem{Hutter}
M.\ Hutter, \emph{Universal Artificial Intelligence}, Springer, 2005.

\bibitem{GameOfLife}
Picture from \url{https://en.wikipedia.org/wiki/Conway\%27s_Game_of_Life}, accessed January 19, 2018. Licensed under the Creative Commons Attribution-Share Alike 3.0 Unported license.

\bibitem{Wheeler}
J.\ A.\ Wheeler, \emph{Law Without Law}, in ``Quantum Theory and Measurement'', ed.\ J.\ A.\ Wheeler and W.\ H.\ Zurek, Princeton Series in Physics, Princeton University Press, 1983.



\end{thebibliography}
\end{document}